\documentclass{pasj00}
\usepackage{color}
\draft

\begin{document}
\SetRunningHead{Ishihara et al.}{AKARI observations of IC4954/4955}
\Received{2007/03/22}
%\Accepted{2007/03/20}

%\title{Nature and Origin of a Young Galactic Open Cluster\\
%(Roslund 4 with reflection nebulae IC4954 and IC4955)\\
%revealed by AKARI First-light Observations}
\title{AKARI infrared imaging of reflection nebulae IC4954 and IC4955}

\author{
  Daisuke \textsc{Ishihara}\altaffilmark{1,*},
  Takashi \textsc{Onaka}\altaffilmark{1},
  Hidehiro \textsc{Kaneda}\altaffilmark{2},
  Toyoaki \textsc{Suzuki}\altaffilmark{2},\\
  Hirokazu \textsc{Kataza}\altaffilmark{2},
  Itsuki \textsc{Sakon}\altaffilmark{1}, 
  Yoko \textsc{Okada}\altaffilmark{2},\\
  Yasuo \textsc{Doi}\altaffilmark{3},
  Naofumi \textsc{Fujishiro}\altaffilmark{2},
  Hideaki \textsc{Fujiwara}\altaffilmark{1},
  Yoshifusa \textsc{Ita}\altaffilmark{2},
  Tuneo \textsc{Kii}\altaffilmark{2}, \\
  Woojung \textsc{Kim}\altaffilmark{2},
  Sin'itirou \textsc{Makiuti}\altaffilmark{2},
  Toshio \textsc{Matsumoto}\altaffilmark{2},
  Hideo \textsc{Matsuhara}\altaffilmark{2},\\
  Hiroshi \textsc{Murakami}\altaffilmark{2},
  Takao \textsc{Nakagawa}\altaffilmark{2},
  Youichi \textsc{Ohyama}\altaffilmark{2},
  Shinki \textsc{Oyabu}\altaffilmark{2},\\
%  Serjeant \textsc{Stephen}\altaffilmark{4},
  Stephen \textsc{Serjeant}\altaffilmark{4},
  Hiroshi \textsc{Shibai}\altaffilmark{5},
  Toshinobu \textsc{Takagi}\altaffilmark{2},
  Toshihiko \textsc{Tanab\'{e}}\altaffilmark{6},\\
  Kazunori \textsc{Uemizu}\altaffilmark{1},
  Munetaka \textsc{Ueno}\altaffilmark{3},
  Fumihiko \textsc{Usui}\altaffilmark{2},
  Takehiko \textsc{Wada}\altaffilmark{2},
\and
  Hidenori \textsc{Watarai}\altaffilmark{7}
}
\altaffiltext{*}{e-mail: ishihara@astron.s.u-tokyo.ac.jp}
\altaffiltext{1}{Department of Astronomy, Graduate School of Science, The University of Tokyo,\\Bunkyo-ku, Tokyo 113-0033, Japan}
\altaffiltext{2}{Institute of Space and Astronautical Science, Japan Aerospace Exploration Agency, \\3-1-1 Yoshinodai, Sagamihara, Kanagawa 229-8510, Japan}
\altaffiltext{3}{Department of Earth Science and Astronomy, University of Tokyo, 3--8--1 Komaba,\\Megro-ku, Tokyo 153--8902, Japan}
\altaffiltext{4}{Astrophysics Group, Department of Physics and Astronomy, \\The Open University, Milton Keynes MK7 6AA}
\altaffiltext{5}{Graduate School of Science, Nagoya University,\\Furu-cho, Chikusa-ku, Nagoya 464--8602, Japan}
\altaffiltext{6}{Institute of Astronomy, University of Tokyo, 2-21-1 Osawa, Mitaka, Tokyo 181-0015, Japan}
\altaffiltext{7}{Office of Space Applications, Japan Aerospace Exploration Agency,\\Tsukuba, Ibaraki, 305-8505, Japan}

\maketitle

\begin{abstract}
We present the observations of the reflection nebulae IC4954 and IC4955 region
with the Infrared Camera (IRC) and the Far-Infrared Surveyor (FIS)
on board the infrared astronomical satellite AKARI
during its performance verification phase.
%The nebulae are associated with the young open cluster Roslund 4
%whose heliocentric distance and age are 2--3 kpc and 4--15 Myr.
We obtained 7 band images from 7 to 160\,$\mu$m with higher spatial
resolution and higher sensitivities than previous observations.  Among the
7 band images, the S11 (11\,$\mu$m) data provide the highest spatial-resolution 
and most sensitive image, in which 10 point-like sources are newly detected in 
the mid-infrared and IC4955 is resolved into two condensations. 
The mid-infrared images show distinct arc-like structures,
inside of which bright sources in the L18W image are located.
The mid-infrared color of the S9W (9\,$\mu$m) and L18W (18\,$\mu$m) bands shows 
a systematic variation around the exciting sources.   
It is red in the
vicinity of the exciting sources, whereas it becomes blue in the surrounding
regions.  This can be interpreted in terms of the decreasing contribution of
thermal emission from dust grains in equilibrium with the
incident radiation field as the distance from the exciting
source increases.  
%The color variation clearly indicates the location of
%exciting stars and on-going star-formation sites, 
The spatial variation in the mid-infrared color 
suggests that the star-formation in IC4954/4955 is
progressing from south-west to north-east.  The FIS data also clearly
resolve two nebulae for the first time in the far-infrared.
The FIS 4-band data from 65\,$\mu$m to 160\,$\mu$m allow us 
to correctly estimate the total infrared luminosity
from the region, which is about one sixth of the energy emitted from the 
existing
stellar sources.  There is little possibility for the presence of
embedded massive stars that have escaped detection.
%The 140 and 160\,$\mu$m images also reveal the existence of high 
%density and cold regions between IC4954 and IC4955, 
%which is supported by optical images.
Five candidates for young stellar objects have been detected as point sources
for the first time in the 11\,$\mu$m image.  
They are located in the red S9W to L18W color regions, suggesting
that current star-formation has been triggered by previous
star-formation activities.
A wide area map of the size of
about $1^\circ \times 1^\circ$ around the IC4954/4955 region
was created from the AKARI mid-infrared all-sky survey data.
Together with the H{\footnotesize I} 21\,cm data, it
suggests a large hollow structure of a degree scale, 
on whose edge the IC4954/4955 region
has been created, indicating 
star formation over
three generations in largely different spatial scales.
\end{abstract}

\section{Introduction}

The reflection nebulae IC4954 and IC4955 are located in 
the Vulpecula constellation
on the Galactic plane around
($l, b$)=($66.96^\circ, -1.26^\circ$).
These nebulae are associated with the young open cluster Roslund 4
\citep{roslund60}.
The heliocentric distance and the age of this cluster are estimated 
based on the analysis of isochrones as
10\,Myr and 2.9\,kpc by \citet{R96},
4\,Myr and 2\,kpc by \citet{P03} (hereafter P03), and
15\,Myr and 2\,kpc by \citet{D04} (hereafter D04).  These studies
suggest the presence of stars of a relatively wide age range
and on-going star-formation in the IC4954/4955 region.  In this paper
we assume the distance to be 2\,kpc.
P03 and D04 also detected three active regions in [SII] images
and some of them are classified as Herbig Haro objects.
The H$\alpha$, [NII] and [SII] emission lines from the cluster members
are detected in optical spectroscopic observations (D04).
This region is an interesting place for the study of star-formation
process because of the co-existence of both 
young and relatively old populations (P03; D04).
Except for the dedicated observation of D04, 
the parallax, radial velocity, proper motions and spectral type
of the cluster members (or clues for the identification of the members) are 
largely unavailable
in existing catalogs because of its far heliocentric distance.

$^{12}$CO observations show that at least one cloud of 10$^5$ $M_\odot$
is associated with this region \citep{LZ}. 
In the infrared,
these nebulae are observed in the IRAS and MSX Galactic surveys.
The IRAS Point Source Catalog (PSC) has entries of three point sources in this 
region,
whereas the MSX map shows distinguished structures in 
four mid-infrared bands of A(8.28\,$\mu$m), C(11.2\,$\mu$m), D(14.3\,$\mu$m) and 
E(21.34\,$\mu$m).
However, the poor spatial resolution of IRAS and the insufficient sensitivity
of the MSX 21\,$\mu$m prevent us from
making detailed analysis of this interesting region in the infrared.
This region is not included in 
the Galactic Legacy Infrared Mid-Plane Survey Extraordinaire
with {\it Spitzer} (GLIMPSE; \cite{glimpse}).

AKARI is the first Japanese infrared astronomical satellite
dedicated for infrared astronomy \citep{murakami7}.
It has two scientific instruments: the Infrared Camera (IRC; \cite{onaka7})
that covers near- and mid-infrared wavelengths of 2--26\,$\mu$m,
and the Far-Infrared Surveyor (FIS; \cite{kawada7}) that covers the
far-infrared spectral range of 50--200\,$\mu$m.
The performance of both instruments has been confirmed by observations
during the performance verification (PV) phase from 2006 April 24 to May 8.
The present paper reports the results of observations of the IC4954/4955
region with both IRC and FIS taken mainly during the PV phase.

The observations and data reduction are described in \S2.
The observational results are shown in \S3.
The nature and the origin of the Roslund 4 region are discussed in \S~4.
Finally, we summarize the results in \S5.

\section{Observations and Data reductions}

AKARI seven band images were taken by two scientific instruments
with three kinds of operation mode.
The observational parameters are summarized in Table \ref{table:data}.
% Details of the observations and the data reduction are described in this section.

\begin{longtable}{rrrrrccc}
\caption{Observational parameters for IC4954/4955 by AKARI}\label{table:data}
\hline\hline
$\lambda_{\rm ref}$&$\Delta\lambda$&FWHM&Pix scale&Scan speed &Instrument&Filter&AOT\\
 ($\mu$m)  & ($\mu$m) & ($''$) & ($''\times''$)  &($''$ s$^{-1}$)&     &      &      \\
\endfirsthead
\endhead
\endfoot
\endlastfoot
\hline
    9      &   4.10   &  5.5    &9.36$\times$9.26  &   215    & IRC &S9W   & ASS  \\
   11      &   4.12   &  4.8    &1.23$\times$2.34  &    30    & IRC &S11   & IRC51\\
   18      &   9.97   &  5.7    &9.36$\times$9.36  &   215    & IRC &L18W  & ASS  \\\hline
   65      &  21.7    &   37    &26.8$\times$26.8  &    15    & FIS &N60   & FIS01\\
   90      &  37.9    &   39    &26.8$\times$26.8  &    15    & FIS &WIDE-S& FIS01\\
  140      &  52.4    &   58    &44.2$\times$44.2  &    15    & FIS &WIDE-L& FIS01\\
  160      &  34.1    &   61    &44.2$\times$44.2  &    15    & FIS &N160  & FIS01\\\hline
\end{longtable}

\subsection{IRC All-Sky Survey (9 and 18 $\mu$m)}
%  J : 1.235 um  1594  Jy
%  H : 1.662 um  1024  Jy
%  K : 2.159 um   666.7Jy

The S9W (9\,$\mu$m) and L18W (18\,$\mu$m) data of the IC4954/4955 region 
were taken
as part of the all-sky survey observations.  Both data were taken simultaneously
with different channels of the IRC, 
MIR-S and MIR-L, which observe sky positions 
separated by about 20$^\prime$ in the cross-scan direction \citep{onaka7}.
In the all-sky survey, the IRC is operated in the scan mode
\citep{ishiharaP}, in which the data of only two rows in the
detector arrays are taken
with the cross-scan width of about 10$^\prime$.  The scan speed is about
215$^{\prime\prime}$\,s$^{-1}$.
The output signals of every four adjacent pixels are binned together to reduce 
the data down-link rate and
the virtual pixel scale is $9.^{\prime\prime}36\times9.^{\prime\prime}36$
in the survey mode.
The seconds confirmation is enabled by the independent data sets taken by
the two separated rows, which allow to efficiently reject high-energy
ionization particle events and largely improves the reliability of
the source detection.
Finer spatial information 
as well as improved sensitivity can be expected in the data reduction 
process 
because two rows are sampled and binned in an interlaced manner
and the observed strip is shifted by 4$^\prime$ in adjacent scan paths
with a given region being observed more than twice on average.
The reset interval
is set as 13.5 s, which corresponds to 306 samplings or 
48.$^{\prime}$9 in the scanning direction.
The length of a reset in the all-sky survey is set as
2.2\,ms (hereafter long reset).
The IC4954/4955 region was observed during 2006 
May 1--8 in the PV phase with 
the descending path of $d\beta/dt<0$
and 2006 November 1--8 with the ascending path of $d\beta/dt>0$, where
$\beta$ is the ecliptic latitude.  Both data are averaged with median in 
the present analysis. The difference 
between the two data sets is less than 10\% in the sky brightness.

Fragments of the data sets covering IC4954/4955 are retrieved from the IRC
all-sky survey database
and reduced automatically by the pipeline software to create a map
\citep{ishihara7}.
The pipeline process includes
linearity correction,
flat correction,
reset anomaly correction (see below),
%the reduction of the scattered light,
source extraction,
rejection of high-energy particle events by the seconds confirmation,
position determination using identified objects,
coaddition of the images obtained by the two rows,
and map construction.

The flux calibration of point sources in the all-sky survey data
are carried out based on a large number of detections of
hundreds of stars in the standard star networks (\cite{cohen9};
\cite{cohen3}; \cite{ishiharaA}).
The flux accuracy is estimated as $\sim 7\%$ for the S9W band
and $\sim 15\%$ for the L18W band for point sources
at the present calibration stage.
The detection limits (5$\sigma$) for point sources are estimated as 
50\,mJy for the S9W band and 120\,mJy 
for the L18W  band.
The calibration for diffuse emission is on-going and 
a relative error of 30\% is assigned at present.
%The same calibration is applied for the flux estimation of the
%nebulae.

Positions of the detected objects were first estimated
based on the data of the gyroscopes and the star trackers
on board and then refined
by using the positions of bright (K$<$8\,mag) 2MASS PSC sources.
The position accuracy is estimated to be 5$^{\prime\prime}$ at present.

\subsection{IRC Slow Scan (11\,$\mu$m)}

The S11 (11\,$\mu$m) band image was obtained during the PV phase with the IRC 
slow scan mode of the Astronomical Observation Template (AOT) IRC51. 
IRC51 is a set of round-trip scans
designed for mapping of a small area of up to $\sim 10^\prime\times 120^\prime$
in about 15\,min of a pointed observation \citep{murakami7, onaka7}.
A higher sensitivity than in the all-sky survey
and a wider sky coverage than in imaging observations can be achieved in this 
observation mode. It also allows to observe very bright sources,
such as IC4954/4955, which will be saturated in the imaging observation.
The scan speed was set as $30^{\prime\prime}$s$^{-1}$ and the S11 filter was 
selected.
The observation of IC4954/4955 was carried out on 2006 May 1.

The focal plane arrays were operated in the same scan mode
as in the all-sky survey observations except that the data binning 
is not made and the full spatial
information is obtained in the cross-scan direction.
Confirmation of source detection was made by round-trip scans.
The grid size of slow scan observations is set as
$1.^{\prime\prime}32 \times 2.^{\prime\prime}34$.
It is determined by the physical pixel size in the cross-scan direction 
($2.^{\prime\prime}34$),
and by the sampling rate and the scan speed
in the scanning direction ($1.^{\prime\prime}32$).
It oversamples the image size of the imaging mode at 11\,$\mu$m 
($\sim$ 4.$^{\prime\prime}$8; \cite{onaka7}).
The time spent for a reset was shortened to 68\,$\mu$s (short reset).
The charge integration curve shows an anomalous behavior 
immediately after a reset (reset anomaly).
The magnitude of this effect depends on the time spent for a reset.
We confirm that the short reset significantly improves the reset anomaly effect.
Gaps in the observed area due to the reset was thus reduced
thanks to the combination of the slower scan speed and the short reset.
The reset period was also shortened to 2.24\,s 
because the gap due to the reset could be ignored.
Consequently the effect of saturation was also reduced.
There is a drift in the offset level owing to the temperature drift 
\citep{ishihara3}.
It is corrected by referring to the signal level during the reset.
%\textcolor[named]{Gray}{
%The signal during the reset corresponds the output with 
%the input shortened and thus indicates the offset level of the
%preamplifier.}
The signal during the reset corresponds to the output with the input 
blocked, and thus indicates the offset level of the preamplifier.
The detection limit is determined by the read-out noise of the detector
rather than the zodiacal background fluctuation
because of the short sampling time.

The data reduction largely follows that for the all-sky survey
except for a few parameters adjusted for the finer pixels,
such as flat correction, and a custom-designed
map reconstruction module is
developed.
At first the data taken in a single scan are arranged
into a $10^\prime\times60^\prime$ strip image by
assuming that the sampling rate and the scan speed of the satellite are 
constant.
Individual sets of images obtained by two separated rows
are aligned with each other and coadded into a combined image.
The position reconstruction of the map is made based on the association with
detected 2MASS sources as in the all-sky survey data.
The positional accuracy of the data
is estimated to be 5$^{\prime\prime}$ at present.

The in-flight calibration of the IRC has been carried 
out for all the imaging
filters in pointed imaging observations based on observations
of the standard star networks \citep{onaka7}.
For the all-sky observations and slow-scan observations, however,
most observations have been made only with the S9W and L18W bands.
The flux calibration for the S11 (11\,$\mu$m) band image in the slow-scan
observations is performed in an indirect way
because no observations of standard stars have been made in the same
scan speed with the same filter. 
First, the calibration for the S9W in the slow-scan mode with
the scan speed of 30\,$''$ s$^{-1}$ is
made by the
comparison of fluxes derived from slow-scan observations with 
those from the all-sky survey for the same stars.
Then the calibration for the S11 band with the scan speed of 30\,$''$ s$^{-1}$
is estimated by
assuming the relative calibration between the S9W and S11 bands
in pointed imaging observations.  Taking account of the uncertainties in
the indirect calibration, we set a conservative uncertainty of 30\% in the
S11 data at present.
%both for point sources and diffuse emission.

\subsection{FIS Slow Scan (65,90,140,160 $\mu$m band)}

FIS observations of IC4954/4955 were executed with N60 (65\,$\mu$m), 
WIDE-S (90\,$\mu$m), WIDE-L 
(140\,$\mu$m), and N160 (160\,$\mu$m) bands in the 2-round-trip 
slow-scan mode (FIS01)
with a scan speed of $15^{\prime\prime}$s$^{-1}$ on 2006 May 3.
The four-band data were taken
simultaneously with correlated double 
sampling (CDS) in about 15\,min of a pointed observation.
Details of the observation 
scheme and the data reduction of the
FIS slow scan data are
described in \citet{ISMkaneda} and \citet{ISMjin}.  % and \citet{ISMmaki}.
At present, the systematic flux
calibration errors for the CDS mode
are estimated to be
30, 40, 50, and 50\% for N60, WIDE-S (90\,$\mu$m),
WIDE-L (140\,$\mu$m), and N160, respectively.
The relative position accuracy among the FIS bands is better than 
10$^{\prime\prime}$ because they were taken simultaneously, but
the absolute accuracy is estimated to be about 1$^\prime$.  
Thus all the FIS images
are aligned relative to the IRC images in the equatorial
coordinates.  
Since the IRC data were taken at a different epoch from
that of FIS, the relative position accuracy is estimated to be
about 30$^{\prime\prime}$ at present.

\section{Results}

\subsection{Infrared images of IC4954 and IC4955}

\begin{figure}[!htb]
\centering
%\FigureFile(165mm,30mm){irc3.eps}
%\FigureFile(130mm,30mm){fis4.eps}
\FigureFile(170mm,30mm){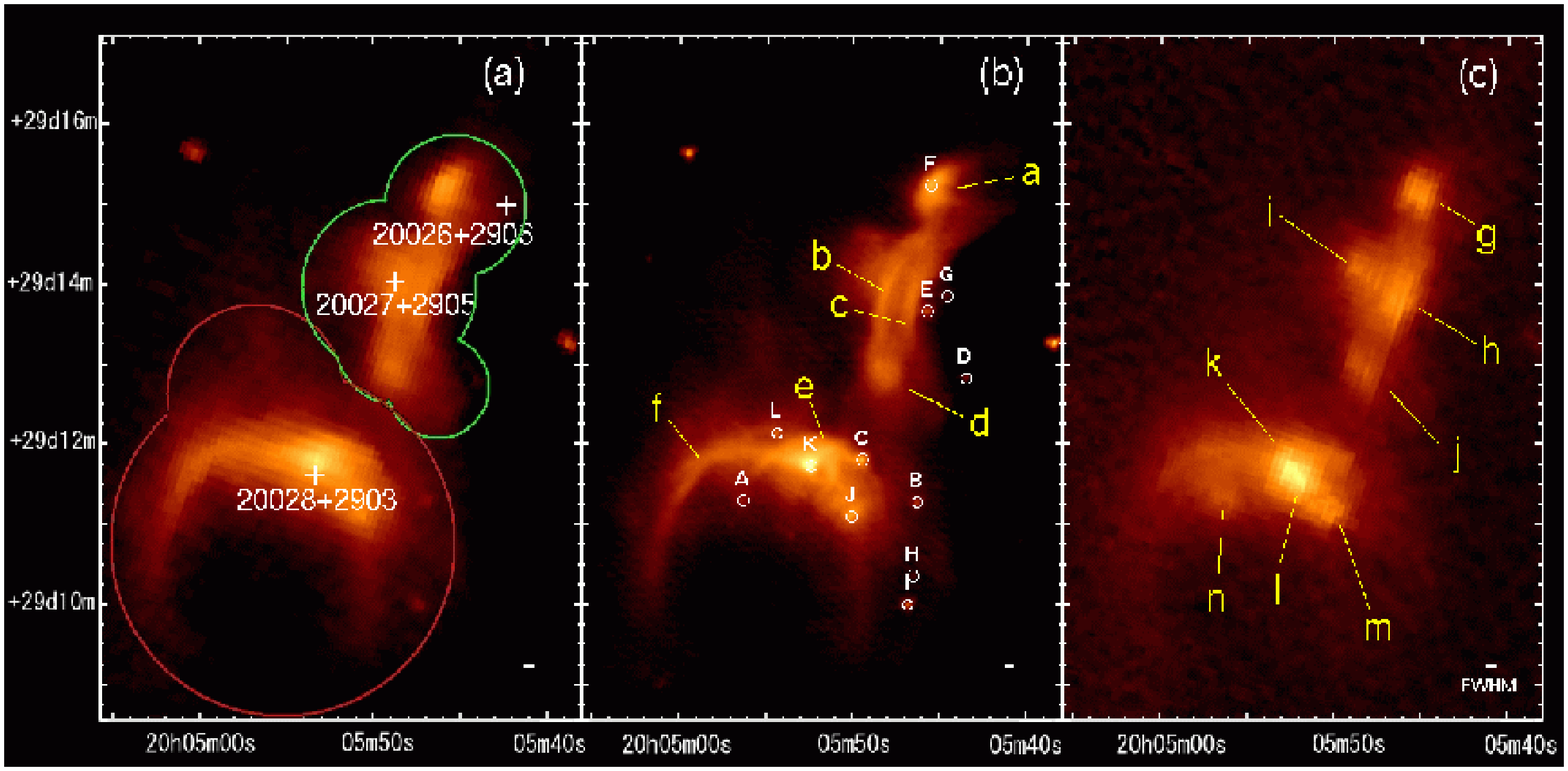}
\FigureFile(130mm,30mm){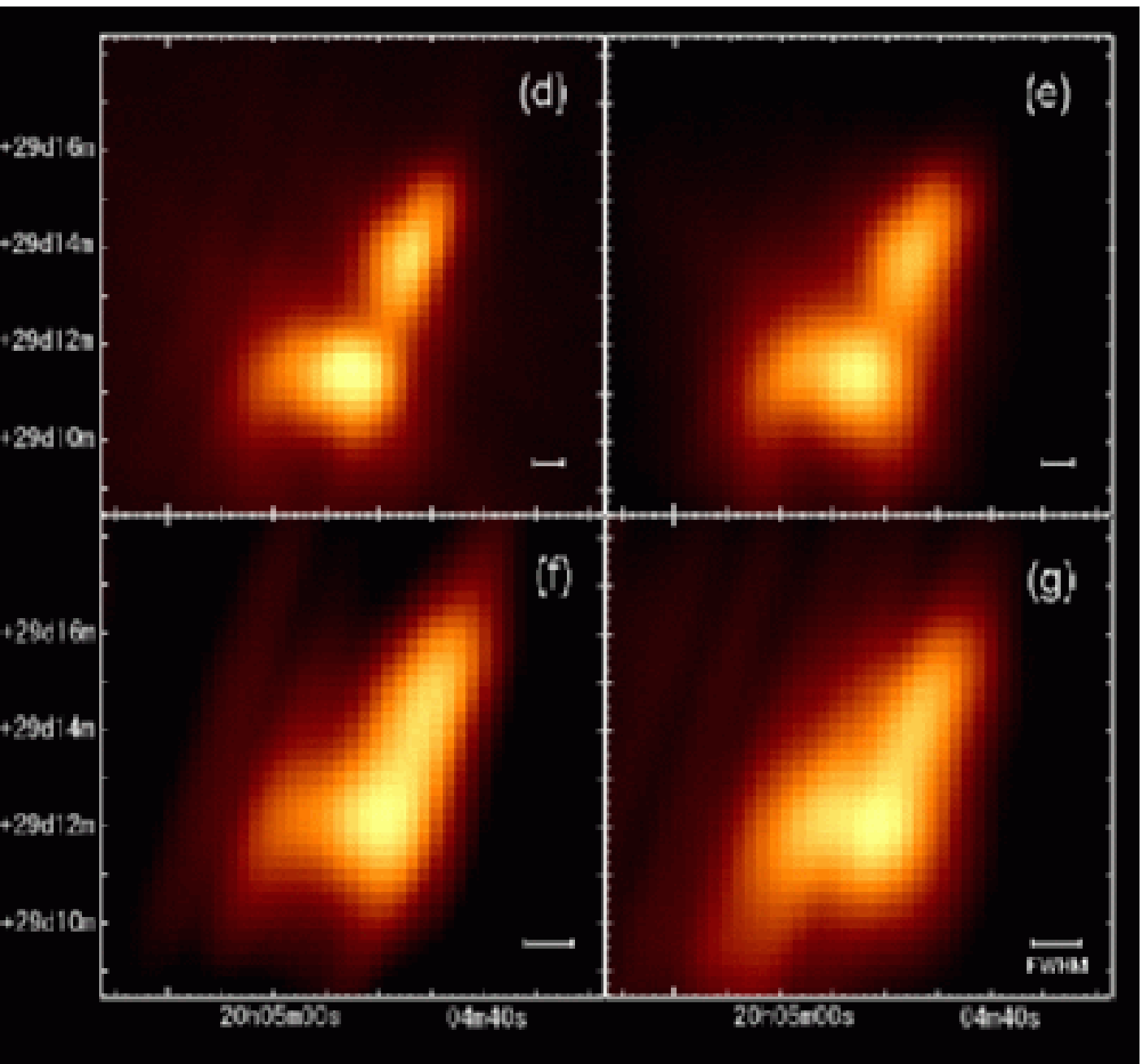}
\caption{IC4954/4955 images by AKARI observations:
(a) S9W (9\,$\mu$m), (b) S11 (11\,$\mu$m), 
(c) L18W (18\,$\mu$m), (d) N60 (65\,$\mu$m), 
(e) WIDE-S (90\,$\mu$m), 
(f) WIDE-L (140\,$\mu$m), and (g) N160 (160\,$\mu$m) bands
in the equatorial (J2000) coordinates.
% The coordinates are shown only in (d) for simplicity.
The integrated areas in the photometry for each nebula are shown on the 
S9W image (a) with the red line for IC4954 and the green line for IC4955.
Three IRAS point sources are also indicated by the plusses.
The symbols $a$--$f$ in the S11 (b) and $g$--$n$ in the L18W (c) indicate 
distinct structures in the images.
The symbols A--L show point-like sources in the S11 image (b).
\label{fig:gallery}}
\end{figure}

The images of IC4954/4955 in AKARI 
S9W, S11, L18W, N60, WIDE-S (90\,$\mu$m), WIDE-L (140\,$\mu$m),
and N160 bands
are shown in figure~\ref{fig:gallery}.
%The optical, mid-infrared and far-infrared three color images of IC4954/4955
%are shown in figure\ref{fig:3color1}.
The northern nebula in the image is IC4955 (enclosed by the green line in 
figure~\ref{fig:gallery}a) 
and the southern nebula is IC4954 (enclosed by the red line).
The AKARI images reveal several distinct
characteristics of the 
nebulae.

The AKARI S9W map by the all-sky survey 
is created with a grid size of $1.^{\prime\prime}56
\times 1.^{\prime\prime}56$ from the virtual pixel size of $9. ^{\prime\prime}36
\times 9.^{\prime\prime}36$.
It is in good agreement with
the MSX A-band (8.28\,$\mu$m) image.
The AKARI S9W band has a better spatial resolution
and the relative spectral response 
similar to that of MSX 
A-band except for the 
inclusion of the 11.2\,$\mu$m unidentified infrared (UIR)
band \citep{onaka7}.
%The MSX A-band image has the grid size of 
%$6^{\prime\prime} \times 6^{\prime\prime}$, created from the detector array
%pixels of $18^{\prime\prime} \times 18^{\prime\prime}$ in size.
%and thus has better spatial resolution than the MSX image.
Three IRAS point sources exist in this region as indicated in the
S9W image.  The location of IRAS~20026+2906 does not match with the
bright spot in the S9W images (source F in figure~\ref{fig:gallery}b)
probably due to the complicated background structures in this region. 

The S11 image has the finest resolution and the deepest sensitivity
among the seven bands owing to its observing mode.
Point-like objects are clearly resolved and denoted by A--L 
(for details see \S 3.2).  Distinct extended structures are denoted
by $a-h$.
The northern nebula IC4955 is spatially resolved
into two condensations ($a$ and $d$) and two arc-like structures ($b$ and $c$).
The two arcs may overlap with each other on the line-of-sight.
Arc-like structures are also found in IC4954 ($e$ and $f$).
All the arcs point toward the north-east of the nebulae and 
extend over the outer edge
of the nebula emission in the optical image.

The AKARI L18W image has a higher signal-to-noise ratio
and higher spatial resolution than the MSX E-band (21.2\,$\mu$m) image.
Point-like objects ($g$--$n$) are clearly seen in figure~\ref{fig:gallery}c.
The spatial distribution of the 18\,$\mu$m emission seems to be
significantly different to that of the S9W or S11 emission.
See \S 4.1 for the discussion on the morphologies of the S9W
and L18W images. 

\begin{figure}[!htb]
\centering
%\FigureFile(80mm,80mm){fig.eps}
%\FigureFile(110mm,110mm){fig.eps}
\FigureFile(110mm,110mm){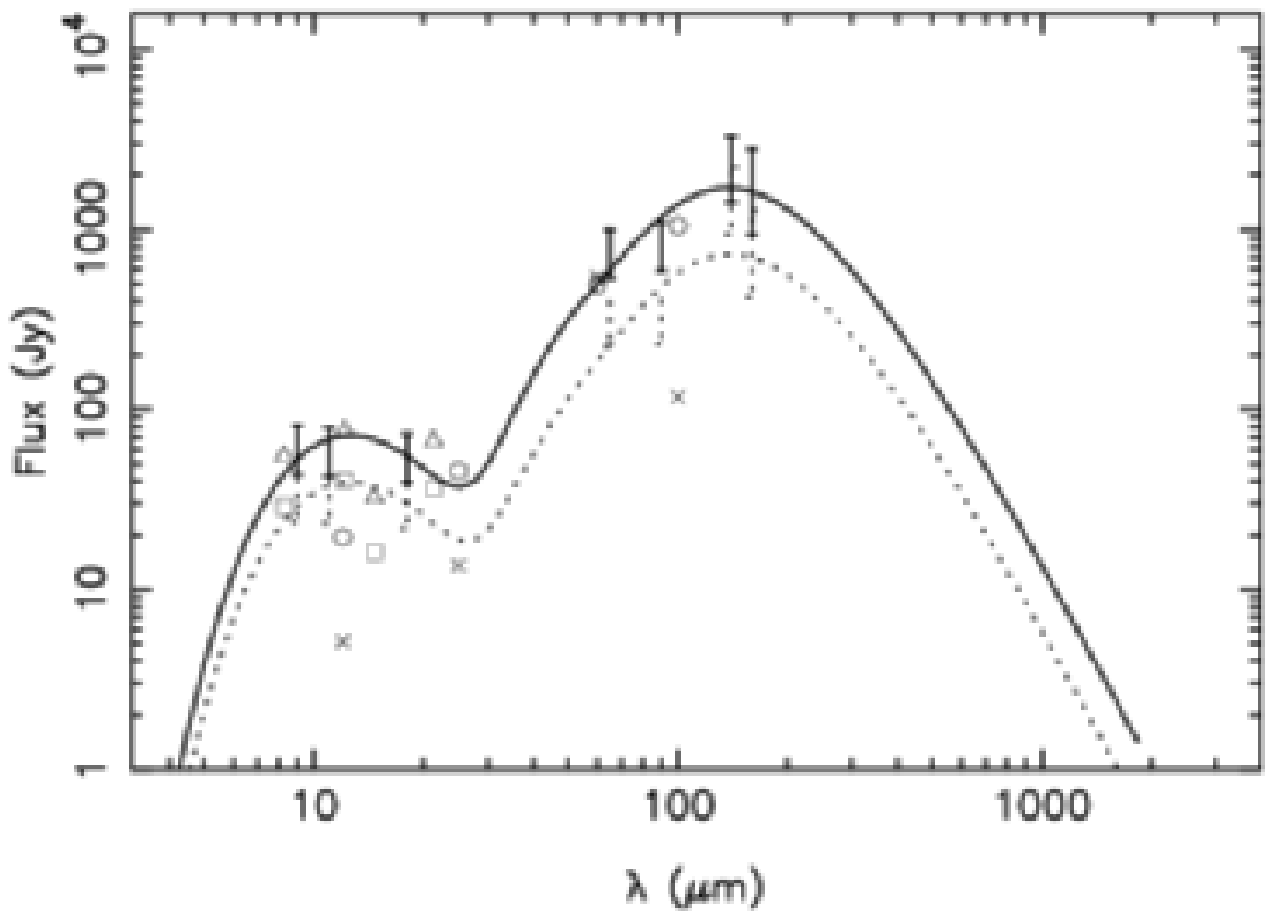}
\caption{
The 9--160\,$\mu$m spectral energy distribution 
of IC4954 and IC4955 measured by AKARI
The solid bars show the observational data for IC4954 and the dotted bars 
for IC4955 derived from the AKARI observations.
The triangles and squares indicate those estimated from
the MSX data for IC4954 and IC4955,
respectively.  The circles and crosses are those from
the IRAS PSC data for IC4954 and IC4955, respectively
(see text for details).  The solid and dotted
curves are fitted 3-temperature dust models (see also text).
\label{fig:SEDs}}
\end{figure}

The AKARI N60 (65\,$\mu$m) and WIDE-S (90\,$\mu$m) image surpass the 
{\it IRAS} 60 or 
100\,$\mu$m images.
The two nebulae are spatially resolved in this wavelength range for 
the first time.
The faint arc-like structure toward the south-east is also detected.
The WIDE-L (140\,$\mu$m) and N160 (160\,$\mu$m) data are unique to AKARI.
These two bands make significant constraints on the dust emission 
(figure~\ref{fig:SEDs}) for the estimate of the total infrared luminosity
(\S~\ref{41}).

The total fluxes from IC4954 and IC4955 are derived from the integrated signals
in the regions enclosed by the red and green lines in figure~\ref{fig:gallery}a
and the sky background is estimated from the surrounding region and subtracted.
The results of IC4954 and IC4955 are summarized in Table \ref{table:sed}
and their spectral energy distributions (SEDs) are plotted in 
figure~\ref{fig:SEDs}.
%The apertures adopted in the photometry are shown in left panel of 
%figure\ref{fig:gallery}.
The uncertainties in Table~\ref{table:sed} include 
the systematic errors.
The MSX image data at band A (8.28\,$\mu$m)
C (12.13\,$\mu$m), D (14.65\,$\mu$m), and E (21.3\,$\mu$m) are also integrated 
over the same regions and the sky background is subtracted similarly.
The results are plotted in figure~\ref{fig:SEDs}.
Except for the band D data, the IRC data are in fair agreement with the MSX 
data, taking account of the uncertainties and the differences in the
band profiles.  The band D data are fainter even compared to other MSX
band data for both nebulae.

In figure~\ref{fig:SEDs} also plotted are the IRAS PSC data.
It is not possible to estimate the fluxes for IC4954 and IC4955 directly from
the IRAS data since the sources are not clearly resolved.
In the figure, the fluxes of IRAS~20028+2903 are plotted as IC4954, whereas the sum 
of the fluxes of
IRAS0026+2906 and IRAS2007+2905 are indicated as IC4955.  Except for the 60\,$\mu$m 
data, the IRAS fluxes are lower than the AKARI/IRC and FIS fluxes and
the agreement is not very good.  The difference can be attributed to the fact 
that 
the IRAS data do not include the diffuse emission correctly in addition to the
position mismatch of IRAS~20026+2906.

\begin{table}
\caption{Photometric results of IC4954 and IC4955$^*$}\label{table:sed}
\begin{center}
\begin{tabular}{cccccccc}
\hline\hline
\multicolumn{3}{l}{AKARI/IRC results}\\
Object & S9W       & S11       &  L18W      & N60       & WIDE-S    & WIDE-L    & N160 \\
       & 9 $\mu$m  & 11 $\mu$m &  18 $\mu$m & 65 $\mu$m & 90 $\mu$m & 140 $\mu$m & 160 $\mu$m \\
\hline
IC4954 &62.2$\pm$18.7 &61.1$\pm$18.3 &56.2$\pm$16.9 &790$\pm$240 &860$\pm$260 &2500$\pm$1000&1890$\pm$570 \\
IC4955 &33.6$\pm$10.1 &33.4$\pm$10.2 &31.2$\pm$9.36 &330$\pm$100 &340$\pm$100 &1700$\pm$680 & 850$\pm$260 \\
\hline
%\end{tabular}
\vspace{-10pt}\\
%\begin{tabular}{ccccc}
\multicolumn{3}{l}{MSX data}\\
name       & 8.28  $\mu$m & 12.13 $\mu$m & 14.65 $\mu$m & 21.34 $\mu$m\\ \hline
IC4954     & 55.15$\pm$0.26 & 75.15$\pm$2.25 & 32.29$\pm$1.29 & 66.84$\pm$4.01 \\
IC4955     & 28.86$\pm$1.44 & 40.25$\pm$1.21 & 16.10$\pm$0.64 & 36.21$\pm$2.17 \\
\hline
%\end{tabular}
\vspace{-10pt}\\
\multicolumn{4}{l}{IRAS PSC sources} \\
IRAS name       & 12  $\mu$m & 25 $\mu$m &  60 $\mu$m & 100 $\mu$m\\ \hline
20026+2906 & 2.741$\pm$0.66  &  $<$0.25       & $<$3.186       & $<$66.76  \\
20027+2905 & 3.356$\pm$0.30  & 14.06$\pm$0.70 & $<$540.3       & $<$50.71  \\
20028+2903 & 20.26$\pm$0.81  & 47.29$\pm$1.89 & 540.3$\pm$48.6  & 1177$\pm$129 \\
\hline
\multicolumn{5}{l}{$^*$ Units are in Jy.}
\end{tabular}
\end{center}
\end{table}

% 11 2.744700e+01 2.990646e-03 IC4954    11 1.502021e+01 1.928618e-03 IC4955
% 20028+2903(4954) 20021+2905(4955)

\subsection{Newly detected point-like sources \label{s3.2}}

\begin{figure}[!htb]
\centering
%\FigureFile(65mm,65mm){newsrc2.eps}
\FigureFile(160mm,160mm){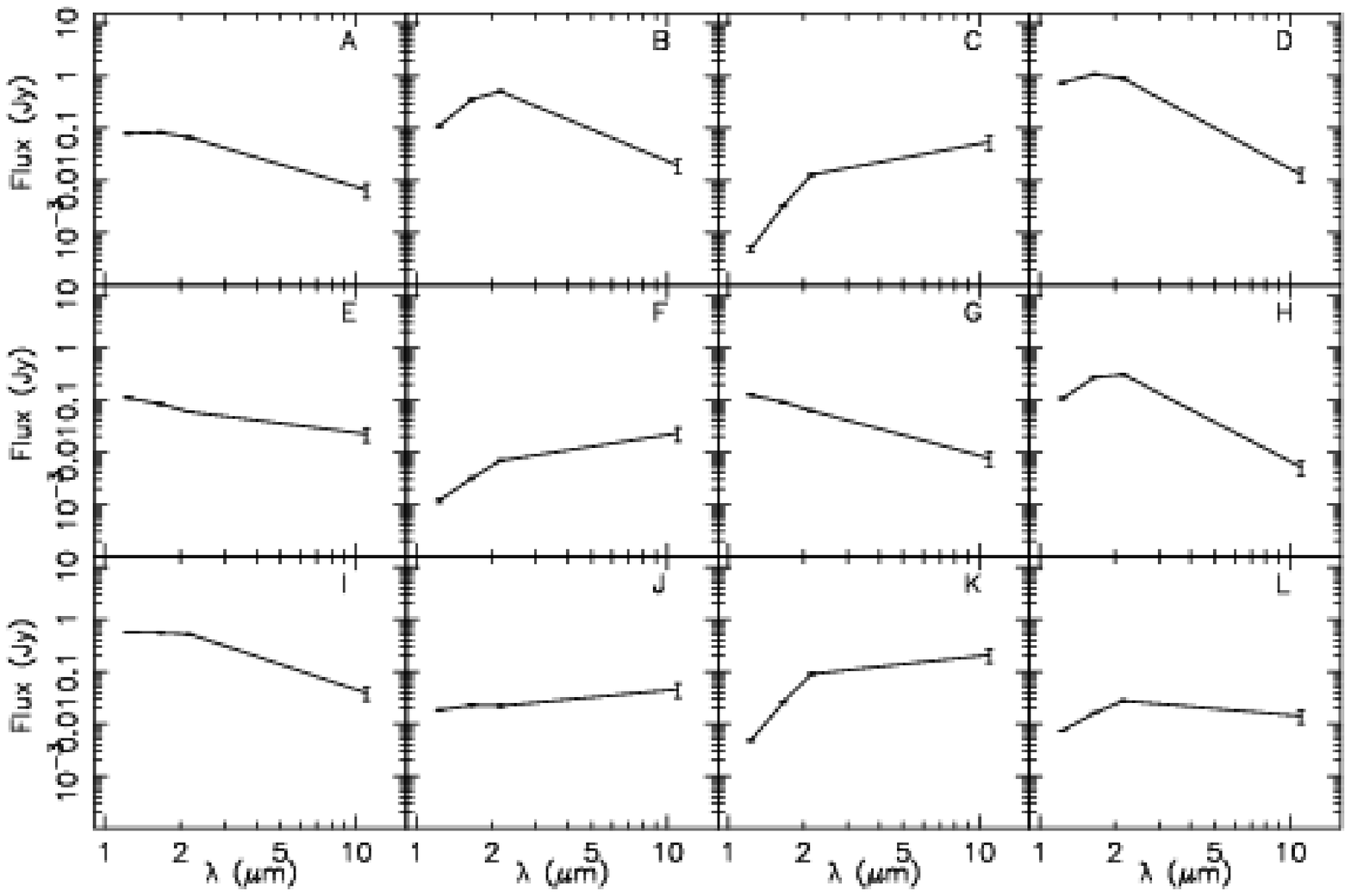}
\caption{
Spectral energy distribution in the near- to mid-infrared 
of 12 detected sources in the S11 image (figure~\ref{fig:gallery}).  
The J, H, and K data are taken from the 2MASS catalog.
%The systematic error for 11\,$\mu$m flux is set as 30\%.
\label{fig:yso}}
\end{figure}

The fluxes and positions of 12 point-like objects detected in the S11 image
(indicated by A--L in figure~\ref{fig:gallery}b)
are summarized in Table~\ref{table:yso}.  
They all have corresponding 2MASS sources as indicated in the table.
The SED of each source is plotted in figure~\ref{fig:yso},
including the 2MASS data.
There is a MSX source (G066.9971-01.2247) close to source F, 
whose C band flux (12.13\,$\mu$m)
is much larger than the S11 flux (see Table~\ref{table:yso}), suggesting
that the MSX flux includes contributions from the surrounding
diffuse emission since the source is located in an extended emission region.  
In the vicinity of source K, there are two
MSX sources (G066.9646-01.2783 and G066.9609-01.2776), whose C
band fluxes are also very large (see Table~\ref{table:yso}).
G066.9646-01.2783 seems to correspond to source K in position
and G066.9609-01.2776 seems to be located close to 
another
optically bright source, 2MASS 20045331+2911469,
around which the S11 image does not detect any point source.
The large MSX fluxes can also be attributed to the diffuse emission
around the source.
Except for F and K, 10 sources are  
detected as point sources for the first time 
in the mid-infrared by the present IRC observations.
%The sources C and F seems to have no optical counter part in DSS image. 
%The sources C, F, J, K, and L show very red SEDs, suggesting that they may
%be embedded young objects.  

\begin{table}[!ht]
\centering
\caption{Point sources detected in the S11 image\label{table:yso}}
%\begin{tabular}{crcrcr}\hline\hline
%   &  (mJy)     &   & (mJy)      &   & (mJy)      \\\hline
%A  &  6.65$\pm$0.33 & E &  22.26$\pm$0.51           & I &   39.25$\pm$0.43          \\
%B  & 19.59$\pm$0.30 & F &  22.96$\pm$0.75           & J &   45.65$\pm$1.10          \\
%C  & 55.63$\pm$1.22 & G &    7.71$\pm$0.18          & K &   208.53$\pm$4.44         \\
%D  & 13.13$\pm$0.30 & H &    5.24$\pm$0.19          & L & %14.26$\pm$0.35         \\\hline
%\begin{tabular}{cccccc}\hline\hline
%ID & R.A. & Decl. & 2MASS ID & Nearest MSX ID & 11\,$\mu$m flux(mJy) \\\hline
%A & 20 04 56 & +29 11 17 & 20045681+2911154 & - &    6.7$\pm$0.3 \\
%B & 20 04 46 & +29 11 16 & 20044653+2911167 & - &   19.6$\pm$0.3 \\
%C & 20 04 49 & +29 11 48 & 20044974+2911486 & - &   55.6$\pm$1.2 \\
%D & 20 04 43 & +29 12 49 & 20044366+2912496 & - &   13.1$\pm$0.3 \\
%E & 20 04 45 & +29 13 39 & 20044585+2913389 & - &   22.3$\pm$0.5 \\
%F & 20 04 45 & +29 15 14 & 20044568+2915152 & G066.9971-01.2247 &   23.0$\pm$0.8 \\
%G & 20 04 44 & +29 13 50 & 20044473+2913512 & - &    7.7$\pm$0.2 \\
%H & 20 04 46 & +29 10 21 & 20044676+2910221 & - &    5.2$\pm$0.2 \\
%I & 20 04 47 & +29 09 59 & 20044710+2910006 & - &   39.2$\pm$0.4 \\
%J & 20 04 50 & +29 11 05 & 20045044+2911060 & - &   45.7$\pm$1.1 \\
%K & 20 04 52 & +29 11 43 & 20045278+2911435 & G066.9646-01.2783 &  208.5$\pm$4.4 \\
%L & 20 04 54 & +29 12 07 & 20045478+2912080 & - &   14.3$\pm$0.3
 %\\\hline
\begin{tabular}{ccccccc}\hline\hline
ID & R.A. & Decl. & 2MASS ID & 11\,$\mu$m flux  
& MSX source & MSX band C flux\\
& \multicolumn{2}{c}{(J2000.0)} &  & (mJy) & &  (Jy)\\
\hline
A & 20 04 56 & +29 11 17 & 20045681+2911154  &    6.7$\pm$0.3 \\
B & 20 04 46 & +29 11 16 & 20044653+2911167  &   19.6$\pm$0.3 \\
C & 20 04 49 & +29 11 48 & 20044974+2911486  &   55.6$\pm$1.2 \\
D & 20 04 43 & +29 12 49 & 20044366+2912496  &   13.1$\pm$0.3 \\
E & 20 04 45 & +29 13 39 & 20044585+2913389  &   22.3$\pm$0.5 \\
F & 20 04 45 & +29 15 14 & 20044568+2915152  &   23.0$\pm$0.8 
& G066.9971-01.2247 & 3.4441\\
G & 20 04 44 & +29 13 50 & 20044473+2913512  &    7.7$\pm$0.2 \\
H & 20 04 46 & +29 10 21 & 20044676+2910221  &    5.2$\pm$0.2 \\
I & 20 04 47 & +29 09 59 & 20044710+2910006  &   39.2$\pm$0.4 \\
J & 20 04 50 & +29 11 05 & 20045044+2911060  &   45.7$\pm$1.1 \\
K & 20 04 52 & +29 11 43 & 20045278+2911435  &  208.5$\pm$4.4 
& G066.9646-01.2783$^*$ & 5.4393\\
  &          &           &                   &    
& G066.9609-01.2776$^\dagger$ & 3.6187 \\
L & 20 04 54 & +29 12 07 & 20045478+2912080  &   14.3$\pm$0.3 \\\hline
\multicolumn{5}{l}{$^*$ Most probable MSX source that corresponds to K.}\\
\multicolumn{5}{l}{$^\dagger$ No corresponding point source in the S11 
image (see text).}
\end{tabular}
\end{table}

\section{Discussion}

\subsection{Infrared characteristics of IC4954 and IC4955 \label{41}}

The total infrared luminosity from the nebular region can be estimated
from the AKARI observations.
The AKARI IRC and FIS data (figure~\ref{fig:SEDs})
are fitted by a three-temperature dust model
with the emissivity proportional to $\lambda^{-2}$.  The three temperatures
are fixed as 20, 40, and 230\,K for both nebulae.  The fitted models are
shown by the solid and dotted lines in figure~\ref{fig:SEDs} for
IC4954 and IC4955, respectively.  The total infrared luminosity $L_{\rm IR}$
is then calculated by

\[
 L_{\rm IR} = \sum_{i=1}^{3} \int{C_i B(T_i,\lambda)\lambda^{-2} d\lambda},
\]
where $T_i = 20, 40,$ and 230\,K for $i=1, 2,$ and 3, respectively,
and $C_i$ is the
fitting constants.  We obtain $L_{\rm IR}= 9.5\times10^{3}$ and
$4.2\times10^{3} L_\odot$ for IC4954 and IC4955, respectively, the sum of the
two nebulae being $1.4\times10^{4} L_\odot$.

The stellar luminosity available for dust heating can be estimated
by summing up the stellar luminosity $L_{\rm *}$
of all the stars located in the region as
\[
 L_{\rm stellar} = \sum_{\rm members}{L_{\rm *}}.
\]

\begin{figure}[!htb]
\centering
%\FigureFile(160mm,30mm){ds7.eps}
%\FigureFile(165mm,30mm){new3color.eps}
%\FigureFile(165mm,30mm){3color_rev1.eps}
\FigureFile(165mm,30mm){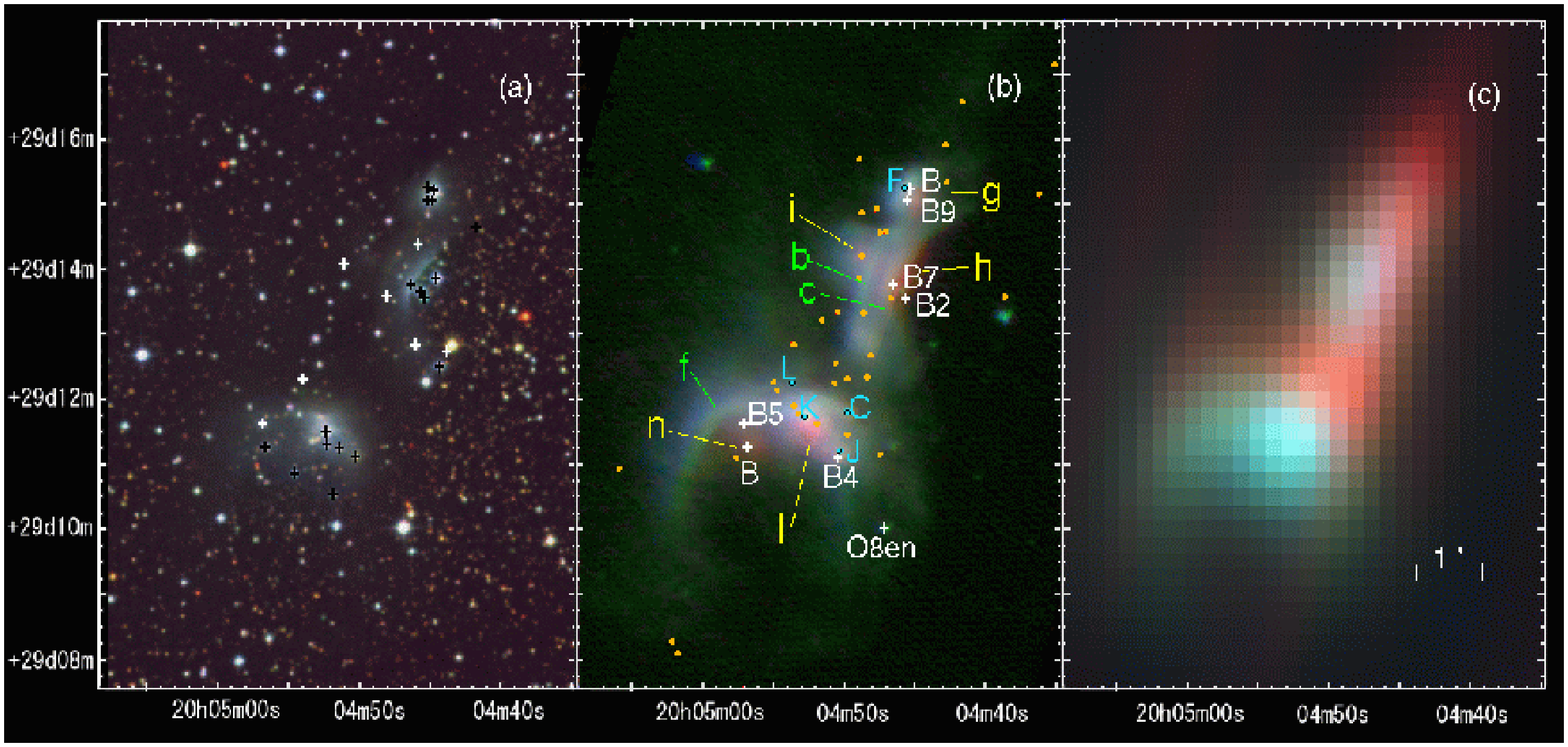}
\caption{Synthesized color images of IC4954/4955 in the equatorial
coordinates: 
(Left) Optical three-color image of B (blue), R (green) and Z (red) 
of the DSS image;
(Middle) mid-infrared three-color image made from the 
S9W (blue), S11 (green) and L18W (red) images taken with the AKARI/IRC.
%The spatial distribution among S9W or S11 and L18W is definitely different.
Optically known early-type stars are indicated by the crosses with the
spectral types \citep{D04, merill}.
The symbols of C, F, J, K, and L in blue show point-like objects in the S11 image,
$a$--$f$ (green) indicate structures seen in the S9W or S11 images,
and $g$--$n$ (yellow) denote structures seen in the L18W image as depicted in
figures~\ref{fig:gallery}b and c.
Red ($H-K>0.9$ and $J-H>0.9$) sources selected from the 2MASS catalog
are shown by the orange circles.
% and other 2MASS sources are shown by the blue circles.
The spectral types of the stars are indicated in white.
(Right) Far-infrared three-color image of IC4954/4955 made from the  
N60 (blue), WIDE-S (90\,$\mu$m) (green) and WIDE-L 
(140\,$\mu$m) (red) images taken with the 
AKARI/FIS.
\label{fig:3color1}}
\end{figure}

The luminosity of each star is estimated simply
based on its spectral type with the assumption that
they are on the main sequence \citep{ZAMS,21,22}.
We select 13 stars that have known spectral types and
are assigned to the member of this region (D04).
We also select other 11 stars whose spectral types are
estimated to be earlier than B7 based on their color.
A total of 24 stars in this region are included in the
estimate of $L_{\rm stellar}$.
%Only stars earlier than B7 are taken into account since the contribution
%from stars of later type is negligible.
%\textcolor[named]{Red}{
%The obvious member stars with spectroscopic
%informations of (D04) are taken into account even if they are late-type.}
The spectral type is taken from the spectroscopic classification
or estimated from the reddening corrected $U-B$ and $B-V$ in D04.
The membership assignment and the UBVRI photometric results 
(D04) were obtained from WEBDA\footnote{http://obswww.unige.ch/webda}.
The location of the 24 stars used in the estimation is shown by the crosses in
figure~\ref{fig:3color1}a.
The O8en star reported in \citet{merill}
located in the south-west of the nebulae (shown in figure~\ref{fig:3color1}b) 
is not included in the estimation
because it seems to be too far from the nebula region.  
Even if we include this star, 
the total stellar luminosity will be increased only by 15\%,
and it will not affect the present conclusion.
The total stellar
luminosity in this region thus estimated is $8.3\times10^{4} L_\odot$.
It is about 6 times of $L_{\rm IR}$.  \citet{leisawitz88} have
indicated that typically only a small fraction ($\sim 0.2$) of the cluster
luminosity is absorbed by dust grains.  The present result is in good
agreement with their results, suggesting that it is not very likely that there
are very luminous stars hidden in dusty environments in this region.

With the dust mass emissivity at 100\,$\mu$m being 0.6\,g\,cm$^{-2}$
\citep{hil83}, the dust mass associated with IC4954 and IC4955 is derived 
to be about
40 and 20 M$_\odot$, respectively.  
The typical size of the nebulae is estimated from the intensity
contours of 10\% of the peak to be
about 1.$^\prime$3 or $2 \times 10^{18}$\,cm.  Assuming that the
gas-to-dust ratio is 100, the average gas density of the infrared
emitting medium is estimated to be about $(2-4) \times 10^5$\,cm$^{-3}$.  
This is in the range of the density of dense photo-dissociation regions
(PDRs), such as the Orion region \citep{tielens85}, indicating
that the IC4954/4955 region is still a young star-forming region
rather than diffuse PDRs, such as the Carina nebula 
\citep{mizutani4}.

\begin{figure}[!htb]
\centering
\FigureFile(165mm,30mm){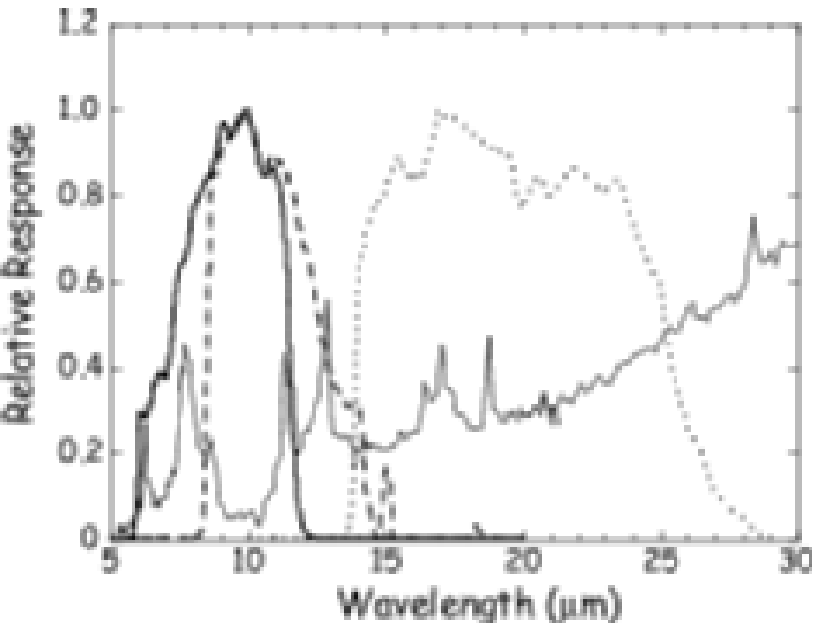}
\caption{
Relative band response of the S9W (thick solid line), 
S11 (thick dashed line), and L18W (thick dotted line) in
units of electron/energy normalized at the peak.  An interstellar
cirrus spectrum at $(l, b) = (355.^\circ 2, 0.^\circ 02)$ taken 
with {\it Spitzer/IRS} is also shown as a reference by the thin
solid line.  It is in units of Jy, but arbitrarily scaled.
\label{fig:band}}
\end{figure}

The color images of IC4954/4955 in the optical, mid-infrared, and far-infrared 
are shown in figure~\ref{fig:3color1}.
In the mid-infrared, the spatial distribution of the 9\,$\mu$m emission 
is similar to that of the 11\,$\mu$m emission,
whereas
the difference in the spatial distribution between S9W and L18W images is 
remarkable. Figure~\ref{fig:band} shows the
relative spectral response of the S9W, S11, and L18W bands.
Also plotted by the thin solid line is an interstellar cirrus
spectrum taken with {\it Spitzer}/IRS for a reference.
The S9W band includes the major UIR bands at 6.2, 7.7, 8.6, and 11.2\,$\mu$m 
except for the 12.7\,$\mu$m band, and is not affected by the continuum
emission longer than 12\,$\mu$m.  The S11 band is affected by only
part of the UIR 7.7\,$\mu$m, but includes the UIR 11.2 and 12.7\,$\mu$m, and
the [NeII]12.8\,$\mu$m line emission.  In the L18W the UIR 17\,$\mu$m complex
is included and the main contributor is the continuum emission longer than
15\,$\mu$m.
No appreciable difference seen
between the S9W and S11 images suggests that 
the spectrum shape between 6 to 13\,$\mu$m
does not vary significantly in the region
and that the line emissions from ionized species is
insignificant compared to the UIR band emission.  
The difference between S9W and L18W
should be related to the variation in the continuum emission longer than
15\,$\mu$m.
The S9W image shows arc-like structures clearly and the L18W image indicates 
high
concentration of the emission in narrow areas.  The difference can be
seen noticeably in the mid-infrared 
color image (figure~\ref{fig:3color1}b).  The emission in L18W is stronger 
inside
the arcs as indicated by red.  Most of the regions 
highlighted
by red color are associated with B-type stars, which are
supposed to play as heating sources of the region: in the 
region $g$, there are a B and a B9 star; the region $h$ is associated with a B2 and 
a B7 star; a B-type star in the region $n$ is suggested to not belong to the member 
of this region on the basis of the
radial velocity (D04) and thus may not be related to this region.  
The other nearby 
B5 star must be a heating source of this region.  The red regions match also 
with bright regions in the H$\alpha$ image (D04).  
We surmise that the red regions are directly heated by B-type stars and
probably associated with ionized gas, whereas the arcs represented by
white color in figure~\ref{fig:3color1}b are PDRs surrounding 
them,
which are characterized by the strong UIR emissions 
in the S9W (cf., \cite{onaka4}).
The strength of the incident
radiation field at the arcs ($b$, $c$, and $f$) is estimated based on
the projected
distance from these heating sources as 900, 3000, and 800, respectively,
in units of the solar neighborhood value ($1.6 \times 10^{-6}$\,W\,m$^{-2}$,
\cite{habing68}).  These are comparable with those of typical PDRs
(cf., \cite{mizutani4}).

IRAS observations indicate the increase of the ratio of 25\,$\mu$m 
to 12\,$\mu$m 
intensities in the vicinity of heating sources for several cases 
\citep{boulanger88, ryter87}.  
The UIR band emission largely contributes to the IRAS 12\,$\mu$m band.
The variation in the intensity ratio has been attributed to the increasing
contribution from thermal emission of large grains rather than the
destruction of the UIR band carriers.  Recent investigations
on the infrared diffuse radiation of our Galaxy and the Large Magellanic
Cloud have in fact shown that the variations in the infrared
SED of the diffuse emission can reasonably be interpreted in terms
of a superposition of the model emissions, in which 
the contribution from dust grains
in radiative equilibrium becomes large in the vicinity of heating sources
together with the effect of destruction
of the UIR band carriers in
strong radiation fields \citep{sakon6, onaka7a}.
The band profile of the IRC S9W is 
shifted to
shorter wavelengths compared to the IRAS 12\,$\mu$m band and thus S9W probes
the UIR band emission more sensitively and is less subject to
thermal emission in longer wavelengths \citep{onaka7}.  
Consequently the effect of thermal emission near the exciting source appears
clearly in the color map of S9W/S11/L18W.  The red color in
figure~\ref{fig:3color1} points to the region strongly heated by
heating sources, whereas the white color indicates the region where the
UIR band emission is dominant.
The observed color variation is well accounted for by 
the increasing contribution from thermal dust emission. 
It further indicates that the UIR-band dominating 
infrared bright regions (white in 
figure~\ref{fig:3color1}b) are always located 
in the north-east side, suggesting the presence of high density regions
in this side.

% YSO
To examine the distribution of young stellar object
(YSO) candidates in this region, objects with red color in the near-infrared are 
selected from
the 2MASS catalog. 
The interstellar reddening to the IC4954/4955 region is estimated
to be $E(B-V)=0.91$ (D04), which corresponds to $(H-K)=0.18$
\citep{rieke85}.
Taking account of the internal extinction, we conservatively 
set the criteria that
$H-K > 0.9$  and $J-H > 0.9$ to select YSO candidates.
They are shown by the red circles and other blue 2MASS sources are indicated
by the blue circles in figure~\ref{fig:3color1}b.
Most of them are located in the white color region as well as in between
the two nebulae, where the optical image shows few stars (see below).   
Only a few red objects are present inside the arcs, indicating 
that YSO candidates are concentrated in dense regions slightly away from those
directly heated by B-type stars.  Five of the 2MASS red objects
are detected in the S11 image (C, F, J, K, and L).  All of them
show YSO-like SEDs in the near- to mid-infrared
(\S~\ref{s3.2}; cf., \cite{whitney3, reach4}), supporting the 
validity of the adopted criteria.
There is also one red object in the relatively red color region
without any corresponding early-type  stars.
It may also play a role of an embedded heating source in the region.
A summary of the relations among the various extended ($a$--$n$) point-like
sources is
given in Table~\ref{table:relation}.

\begin{table}[!ht]
\centering
\caption{Structures and point sources in the mid-infrared of
the IC4954/4955 region\label{table:relation}}
\begin{tabular}{cccl}\hline\hline
\multicolumn{2}{c}{S11} & S18W & \multicolumn{1}{c}{Comments}\\
Extended & Point-like & Extended \\
\hline
$a$  &  F  & $g$ & \hbox{The peak positions of $a$ and $g$ do not coincide. }\\
&&&\hbox{F is located at the peak of $a$.}\\
$b$ & & $i$ & The peak positions do not coincide.  \\
&&&\hbox{There may be a hidden source at $i$.}\\
$c$ & E, G & $h$ & \hbox{The peak positions of $c$ and $h$ do not coincide. }\\
&&&\hbox{E and G are heating sources for $h$.}\\
$d$ & & $j$ & The peak positions are matched within the uncertainties.\\
    & K & $k,l$ & K coincides with $k$.  No counterpart 
    is seen for $l$.\\
    & A & $n$  &  The position of A coincides with $n$.  A is listed as a non-member.\\
    & J & $m$  & J is a heating source for $m$.\\
\hline    
\end{tabular}
\end{table}

% Dark nebulae
Comparison of  figures~\ref{fig:3color1}a and c indicates that
the density of stars is low in between IC4954 and IC4955
in the optical image,
whereas the dust emission is seen in the corresponding region in the WIDE-L (140\,$\mu$m)
image.  The size of the region ($120^{\prime\prime}$)
is sufficiently larger than the FWHM of the beam size
of the WIDE-L band ($\sim58^{\prime\prime}$; \cite{kawada7}), thus
the presence of the far-infrared emission in between the two nebulae 
is not spurious.
It can be surmised that the region is a dark nebula filled with
dust grains.
The region around source F in figure~\ref{fig:gallery}b also
appears dark in the optical image, suggesting that source F is
surrounded by a large amount of dust.

% Summary
The mid-infrared color variation, the location of early-type stars, and 
the distribution of red objects strongly suggest that YSOs are being
born in the region on the
north-east side of the arcs.  Their formation may be
triggered by the existing early-type stars and star-formation
is propagating from south-west to north-east.  
This picture is revealed by the high spatial resolution 
(especially in 11\,$\mu$m) and the multi-band (especially
140 or 160\,$\mu$m) infrared data of AKARI.
Because of the arc-like shape rather than
the cometary shape, it is conjectured that the regions where YSOs are
concentrated in
IC4954/4955 are not pre-existing clouds imploded and/or evaporated
by the effect of stellar winds, such as in the Elephant Trunk Nebula
\citep{reach4}, but are rather triggered by the `collect and collapse' type
mechanism with stellar winds of the heating stars \citep{elm98}.
Rather uniformly distributed YSOs may also support this interpretation
(e.g., \cite{efremov98}).

\subsection{Origin of Roslund 4\label{42}}

\begin{figure}[!htb]
\centering
%\FigureFile(160mm,160mm){widecmp.eps}
\FigureFile(160mm,160mm){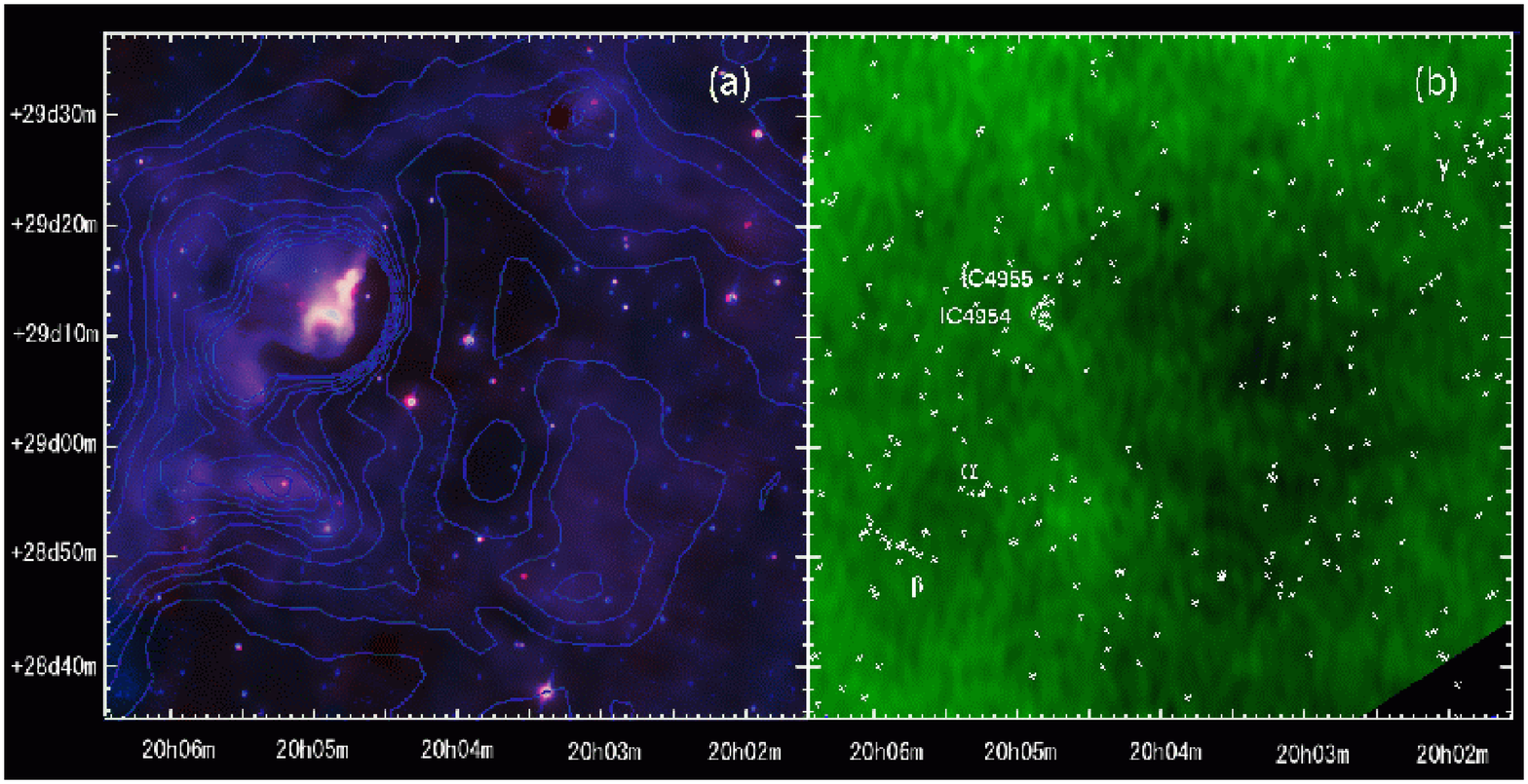}
\caption{
(Left) Color image made from the S9W and L18W all-sky survey
data of AKARI/IRC around IC4954/4955.
The contours indicate the ISSA 100\,$\mu$m intensity.
(Right) CGPS H{\footnotesize I} 21\,cm map integrated over the velocity range of 
0--30\,km\,s$^{-1}$ (see text).
The crosses indicate red sources selected from the 2MASS catalog.
\label{fig:wide}}
\end{figure}

To investigate the origin of the IC4954/4955, mid-infrared maps of
the surrounding region 
of about $1^\circ \times 1^\circ$ at 9 and 18\,$\mu$m
are created from the IRC all-sky survey data.  The S9W/L18W color image
is shown in figure~\ref{fig:wide}a with the contours of the 100\,$\mu$m data
of the IRAS Sky Survey Atlas (ISSA, \cite{wheelock94}).  
It indicates that there is a cavity of 
low mid-infrared emission centered around ($\alpha$, $\delta$)
=(20h 03m, +29$^\circ$.00) and that the IC4954/4955 region
is located on an edge of the cavity.  
%A part of the eastern edge is also recognized 
%in south-east in AKARI WIDE-S \textcolor{red}{(90\,$\mu$m)}
image (figure~\ref{fig:gallery}).

\begin{figure}[!htb]
\centering
%(J\(BFigureFile(90mm,90mm){danmen.eps}
\FigureFile(110mm,110mm){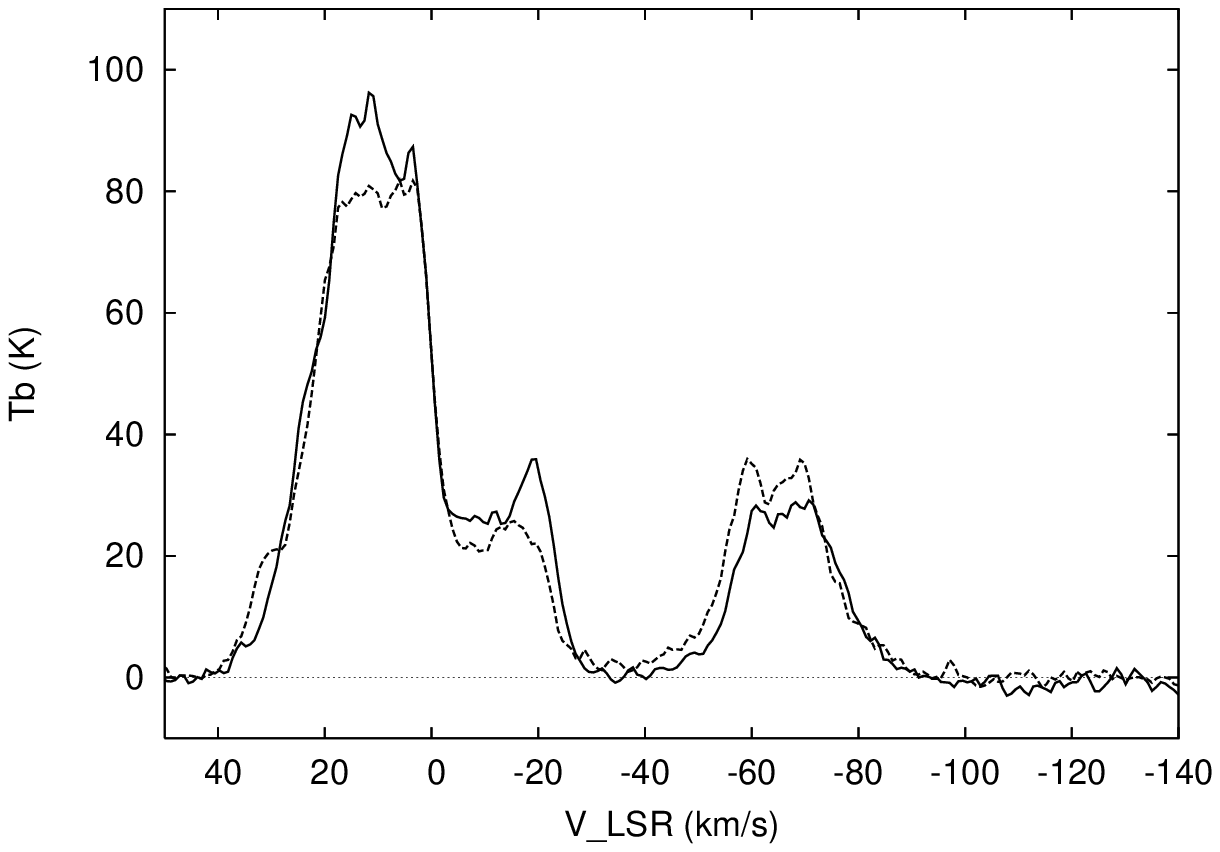}
\caption{
Intensity vs velocity plot of the H{\footnotesize I} 21\,cm 
data (CGPS).  
The solid line indicates at the IC4954/4955 region,
whereas the dashed line corresponds to the center of the cavity
(figure~\ref{fig:wide}).
\label{fig:posdef}}
\end{figure}

To examine and confirm the presence of the
cavity, H{\footnotesize I} 21\,cm data of the region
are obtained from the Canadian Galactic Plane
Survey (CGPS)\footnote{http://www.ras.ucalgary.ca/CGPS} and examined.  
Figure~\ref{fig:posdef} shows a plot of the intensity vs. velocity
of the H{\footnotesize I} data.
It indicates that the H{\footnotesize I} gas in this region has 
a velocity range of 0--30km\,s$^{-1}$.  This is in good agreement of
the velocity range of CO emission associated this region 
of 6--19\,km\,s$^{-1}$ \citep{LZ}.
The H{\footnotesize I} intensity integrated over 0--30km\,s$^{-1}$ is shown in
figure~\ref{fig:wide}b, which clearly supports
the presence of a low density cavity around the center of the map.
In figure~\ref{fig:wide}b also plotted by the crosses are YSO candidates.
They are selected from the 2MASS catalog
based on the same criteria, $H-K > 0.9$ and
$J-H>0.9$, as in figure~\ref{fig:3color1}b.  
In addition a condition of $13.5<K<15.0$ is 
estimated from the K magnitude of the YSO candidates in 
figure~\ref{fig:3color1}b and applied
to exclude foreground and background sources.  
The symbols $\alpha$, $\beta$, and $\gamma$
indicate the regions where concentration of YSOs is seen.
YSO candidates seem to be located on
the edge of the cavity.  
It is most clearly seen in the eastern edge,
which includes the IC4954/4955 region as well as the regions
$\alpha$ and $\beta$, however, only a few YSOs
are found at the north and south edges.

The radius of the cavity is estimated as about 10\,pc.  
The cavity is elongated and partly collapsed.
Such structures are commonly seen in Galactic bubbles \citep{churchwell6}.
The cavity of the IC4954/4955 may have been created by
supernovae (SNe) or O-type stars.   
Progenitor candidates
have been searched for in the 
% ROSAT point source catalog \citep{ROSAT},
Tycho-2 spectral type catalog \citep{Tycho2}, 
All-sky compiled catalog \citep{ASCS},
Mean radial velocities catalog \citep{radial}, 
and SNRs catalog \citep{green}.
but any progenitors of such a kind are not found.
Any SNR like structures are not found either in the soft X-ray maps of
ROSAT or in the NVSS 1.4\,GHz map \citep{NVSS},
although there seem to be some point-like condensations in the region.
% 1RX 200429.9+290215 and 1RX 200240.1+290703,
% It also contains both younger clusters (IC1396, S140 etc.) 
% and older cluster (NGC7129), similar to IC4954/4955.
The cavity of the size of $\sim$40\,pc in IC1396 is made by a 
single O6.5 star \citep{patel5}.  Thus
the O8en star located at the south-west of the IC4954/4955 
(figure~\ref{fig:3color1}b) could be the progenitor
because the dense region is spread at the north-east of the cavity.
However the star 
may be too young if the age of Roslund 4 is 15\,Myr and
the star-formation was delayed relative to the trigger event because
of the internal motions \citep{elm98}.
Since SNRs disappear in the time scale of $\sim$1\,Myr
and the age of the cavity must be older than 4\,Myr
based on the age of the Roslund 4, it is not unexpected that
there is no direct sign for the progenitor.
The Cepheus bubble \citep{patel8} is also thought to
be formed by O-type stars or supernovae, but
shows no direct hint of them.
The present AKARI observations have revealed another bubble of
small scale, which seems to trigger past (4--15\,Myr old) star-formation 
in IC4954/4955.  It also indicates the `current star-formation' in the
region, which is triggered by stars of the second generation.  
This trigger seems to be
different from the `globule-squeezing' type
in the Elephant Trunk Nebula near
the Cepheus bubble.  The present
observations indicate triggered star-formation over the three generations
in a different spatial scale in the IC4954/4955.

\section{Summary}

With the two scientific instruments (IRC and FIS) on board AKARI
we obtained 7 band images of IC4954/4955 
from 7 to 160\,$\mu$m with higher spatial
resolution and higher sensitivities than previous observations.
Based on these observations we obtained the following results.

(1) 
The mid-infrared images at 9, 11, and 18\,$\mu$m
reveal several distinct structures in the region.
Particularly they show three arc-like structures that seem to have been
created by stellar winds from the existing B-type stars.
The difference between the S9W (9\,$\mu$m) and S11 (11\,$\mu$m)
images is not significant, indicating that the variation in the spectrum
of 6--13\,$\mu$m is not large and that 
the contribution of line emission
from ionized gas, such as [NeII]12.8\,$\mu$m, is relatively small compared
to the UIR band emissions at 6.2, 7.7, and 11.2\,$\mu$m.

(2) 
The S9W (9\,$\mu$m) to L18W (18\,$\mu$m) images appear systematically 
different
from each other.  The L18W emission is strong near the exciting stars,
whereas the S9W emission is dominant in the surrounding region.
The S9W filter probes the UIR band emission more effectively than
the IRAS 12\,$\mu$m because its band profile is shifted to
shorter wavelengths.
We interpret the systematic mid-infrared color variation 
in terms of the decreasing contribution of
thermal dust emission with the distance from the exciting
source.  The color variation clearly indicates the location of
exciting stars, suggesting that the star-formation in IC4954/4955 is
progressing from south-west to north-east.

(3)
Young stellar objects are
distinguishable for the first time at 11\,$\mu$m.  
They are located in the large S9W to L18W ratio regions, suggesting
that current star-formation has been triggered by previous
star-formation activities.

(4)
The FIS data allow to correctly estimate the total infrared luminosity
from the region, which is about one sixth of
the energy emitted from the existing early-type stars.
Thus there is little possibility that
embedded luminous stars have escaped detection.
It also suggests that the total dust mass of the infrared emission
is about 60\,$M_\odot$.
The 140 and 160\,$\mu$m images reveal the presence of a high 
density region between IC4954 and IC4955, 
which is also supported by optical images.

(5)
The IRC all-sky survey data 
together with the H{\footnotesize I} 21\,cm data
further suggest the presence of a bubble-like structure of a degree scale, 
on whose edge the IC4954/4955 region
is located, indicating 
triggered star formation over
three generations.

The IC4954/4955 region is not a massive star-formation
region, and is currently populated by B-type stars.  The mass
of the infrared emitting material is also not huge.  The suggested hole
of the interstellar matter is not large and thus should be created by
a less energetic event compared to IC1396.  The present observations
suggest that even in such a moderate star-forming region, the sequential
star-formation occurs and is on-going at present.  
Medium-scale star-formation could be common
and should be studied
in future investigations.

AKARI continues to provide significant 
data for the study of interstellar medium and star forming regions
owing to its high
sensitivity, wide wavelength coverage and wide mapping capability,
during the course of its mission.

AKARI is a JAXA project with the participation of ESA.
We thank all the members of the AKARI project for their continuous help and support.
We thank T. Negishi, I. Maeda, H. Mochizuki, K.-W. Chan, S. Fujita,
C. Ihara, H. Ikeda, T. Yamamuro, and N. Takeyama for their contributions
in the development of the IRC and D. Jennings for providing us with the
calibration sources. We also thank K. Imai, M. Ishigaki, H. Matsumoto,
N. Matsumoto, T. Tange for supporting daily operation of the IRC.
The contribution from the ESAC to the pointing reconstruction is greatly acknowledged.
We also thank M. Cohen for helping us in the flux calibration, 
T. Muller for providing us the 
fluxes of asteroids at the observing times, G. White for careful reading
of the manuscript, and T. L. Roellig for
the IRS spectrum of the cirrus cloud.
We also thank M. Oyadomari, S. Sugiyama, K. Mizushima, T. Tohma,
N. Shigemoto, M. Tomizawa, M. Shinano, M. Suzuki, M. Akutagawa
especially for supporting critical phase of AKARI
as well as supporting long-term pre-flight tests.
We also thank S. Arimura, K. Ito, T. Murakami, % W. Inoue
for supporting daily operation of the AKARI.
We thank H. Iida, K. Higashino, K. Sato, A. Genba, N. Sumi and C. Ihara,
and all the members of the altitude and orbit control team, 
for giving us the advices in the re-construction of the initial data
as well as pre-flight and post-flight daily hard works.
We also thank S. Yoshida, Yu. Ochi, Yo. Ochi, K. Okamoto, S. Tsunematsu,
and all the members of the cryogenic team
for supporting long-term pre-flight cooling tests.
W.K., I.S., and T.T. have been financially supported by the Japan
Society of Promotion of Science (JSPS).
This work is supported in part by a Grant-in-Aid for Scientific Research
on Priority Areas from the Ministry of Education,
Culture, Sports, Science, and Technology of Japan and Grants-in-Aid for
Scientific Research from JSPS.

The research presented in this paper used data from the Canadian
Galactic Plane Survey, a Canadian project with international partners,
supported by the Natural Sciences and Engineering Research Council.
We also made use of the ROSAT Data Archive of the Max-Planck-Institut
f\"ur extraterrestrische Physik (MPE) at Garching, Germany.


\begin{thebibliography}{}
\bibitem[Barbier-Brossat \& Figon (2000)]{radial}
 Barbier-Brossat, M., Figon, P., 2000, \aaps, 142, 217
\bibitem[Benjamin et al.(2003)]{glimpse}
 Benjamin, R., A. et al. 2003, \pasp, 115, 953
\bibitem[Boulanger et al.(1988)]{boulanger88}
 Boulanger, F. et al. 1988, \apj, 332, 328
\bibitem[Churchwell et al.(2006)]{churchwell6}
 Churchwell, E., et al. 2006, \apj, 649, 759
\bibitem[Clarke et al.(2005)]{clarke}
 Clarke, A. J., Oudmaijer R. D. and  S. L. Lumsden, 2005, \mnras, 363, 1111
\bibitem[Cohen et al.(1999)]{cohen9}
 Cohen, M., et al. 1999, \aj, 117, 1864
\bibitem[Cohen et al.(2003)]{cohen3}
 Cohen, M., Megeath, S. T., Hammersley. P. L., Martin-Luis, F., \& Stauffer, J., 2003, \aj, 125, 2645
\bibitem[Condon et al.(1998)]{NVSS}
Condon, J. J., Cotton, W. D., Greisen, E. W., Yin, Q. F., Perley, R. A., Taylor, G. B., \& Broderick, J. J. 1998, \aj, 115, 1693
\bibitem[Cot\'{e} \& van Kerkwijk(1993)]{cote}
 Cot\'{e}, J., \& van Kerkwijk, M. H., 1993, \aap, 274, 870
\bibitem[De Jager et al.(1987)]{22}
 De Jager, C., \& Nieuwenhuijzen, H, 1987, \aap, 177, 217
\bibitem[Delgado et al.(2004)]{D04}
 Delgado, A. J., Miranda, L. F., Fernandez, M., Alfaro, E. J., 
 2004, \aj, 128, 330
\bibitem[Efremov \& Elmegreen(1998)]{efremov98}
 Efremov, Y. N., \& Elmegreen, B. G.\ 1998, \mnras, 299, 643
\bibitem[Elmegreen(1998)]{elm98}
 Elmegreen, B. G.\ 1998, in Origins, ed. C. E. Woodward, J. M. Shull,
 \& H. A. Thronson, Jr., ASP Conf. ser. 148, 150
\bibitem[Green et al.(2006)]{green}
Green, D. A., 2006, {\it A Catalogue of Galactic Supernova Remnants (2006 Aprill version)}
Laboratory, Cambridge, United Kingdom (available at "http://www.mrao.cam.ac.uk/surveys/snrs/").
\bibitem[Habing(1968)]{habing68}
 Habing, H. J.\ 1968, \bain, 19, 421
\bibitem[Hildebrand(1983)]{hil83}
 Hildebrand, R. H.\ 1983, \qjras, 24, 267
\bibitem[Ishihara et al.(2003)]{ishihara3}
 Ishihara, D., et al. 2003, Proc. of SPIE, 4850, 1008
\bibitem[Ishihara et al.(2006a)]{ishiharaP}
 Ishihara, D., et al. 2006a, \pasp, 118, 324
\bibitem[Ishihara et al.(2006b)]{ishiharaA}
 Ishihara, D., et al. 2006b, \aj, 131, 1074
\bibitem[Ishihara et al.(2007)]{ishihara7}
 Ishihara, D., et al. 2007, in prep.
\bibitem[Johnson(1996)]{21}
 Johnson, H. L., 1966, \araa, 4, 193
%\bibitem[Kaneda et al.(2007a)]{kaneda7}
% Kaneda, H., et al. 2007a, \pasj, this volume
\bibitem[Kaneda et al.(2007b)]{ISMkaneda}
 Kaneda, H., et al. 2007b, \pasj, this volume, \#3067
\bibitem[Kawada et al.(2007)]{kawada7}
 Kawada, M., et al. 2007, \pasj, submitted
\bibitem[Kharchenko (2001)]{ASCS}
 Kharchenko, N. V., 2001, {\it Kinematika i Fizika Nebesnykh Tel}, 17, 5, 409
\bibitem[Kerton(2002)]{kerton2}
 Kerton, C. R., 2002,\aj, 124, 3449
\bibitem[Leisawitz \& Hauser(1988)]{leisawitz88}
 Leisawitz, D., \& Hauser, M. G. 1988, \apj, 332, 954
\bibitem[Leisawitz et al.(1989)]{LZ}
 Leisawitz, D., Bash, F. N., and Thaddeus P., 1989, \apjs, 1989, 70, 731
\bibitem[Makiuti et al.(2007)]{ISMmaki}
 Makiuti, S., 2007, \pasj, submitted
\bibitem[Merrill et al.(1942)]{merill}
 Merrill, P., W., Burwell, C., G., \& Miller, W., C., 1942, \apj, 96, 15
\bibitem[Mizutani et al.(2004)]{mizutani4}
 Mizutani, M., Onaka, T., \& Shibai, H.\ 2004, \aap, 324, 579
\bibitem[Murakami et al.(2007)]{murakami7}
 Murakami, H., et al. 2007, \pasj, submitted
\bibitem[Nakagawa et al.(2007)]{nakagawa7}
 Nakagawa, T., et al. 2007, \pasj, submitted
\bibitem[Neugebauer et al.(1984)]{iras}
 Neugebauer, G., et al. 1984, \apjl, 278, L1
\bibitem[Onaka(2004)]{onaka4}
 Onaka, T.\ 2004, in Astrophysics of Dust, ASP Conf. ser. 309, 163
\bibitem[Onaka et al.(2007a)]{onaka7a}
 Onaka, T., Tokura, D., Sakon, I., Tajiri, Y. Y., Takagi, T. \& Shibai, H.
 2007a, \apj, 654, 844
\bibitem[Onaka et al.(2007b)]{onaka7}
 Onaka, T., et al. 2007b, \pasj, this volume
\bibitem[Oudmaijer, R. D. and Dolf de Winter(1995)]{oudmaijer}
 Oudmaijer, R. D. \& Dolf de Winter, 1995, \aap, 295, L43
\bibitem[Patel et al.(1998)]{patel8}
 Patel, N. A., Goldsmith P. F., Heyer M. H., Snell R. L., \& 
 Pratap P., 1998, \apj, 507, 241 
\bibitem[Patel et al.(1995)]{patel5}
 Patel, N. A., Goldsmith P. F., Snell R. L., Hezel T., \& Xie T., 
 1995, \apj, 447, 721
\bibitem[Phelps(2003)]{P03}
 Phelps R. L., 2003, \aj, 126, 826
\bibitem[Price et al.(2001)]{MSX}
 Price, S. D., et al. 2001, \aj, 121, 2819
\bibitem[Racine(1996)]{R96}
 Racine, R.\ 1969, \aj, 74, 816
\bibitem[Reach et al.(2004)]{reach4}
 Reach, W. T., et al. 2004, \apjs, 154, 385
\bibitem[Rieke \& Lebofsky(1985)]{rieke85}
 Rieke, G. H., \& Lebofsky, M. J.\ 1985, \apj, 288, 618
\bibitem[Roslund(1960)]{roslund60}
 Roslund, C.\ 1960, \pasp, 72, 205
\bibitem[Ryter et al.(1987)]{ryter87}
  Ryter, C., Puget, J. L., M\'{e}rault, M., 1987, \aap, 186, 312
\bibitem[Sakon et al.(2006)]{sakon6}
 Sakon, I., et al. 2006, \apj, 651, 174
%\bibitem[Sakon et al.(2007)]{sakon7}
% Sakon, I., et al. 2007, \pasj, submitted
\bibitem[Schmidt-Kaler(1982)]{ZAMS}
 Schmidt-Kaler, Th. 1982,
{\it Landolt-B\"{o}rnstein: Numerical Data and Functional Relationships in
Science and Technology}, edited by K. Schaifers and H.H. Voigt
(Springer-Verlag, Berlin), VI/2b
\bibitem[Suzuki et al.(2007)]{ISMjin}
 Suzuki, J., 2007, \pasj, submitted
\bibitem[Taylor et al.(1996)]{radio}
 Taylor, A. R., Goss, W. M., Coleman, P. H., van Leeuwen, J., \& Wallace, B. J., 2003, \aj, 125, 3145
\bibitem[Taylor et al.(2003)]{cgps}
 Taylor, A. R., 2003, \aj, 125, 3145
\bibitem[Tielens \& Hollenbach(1985)]{tielens85}
 Tielens, A. G. G. M., \& Hollenbach, D.\ 1985, \apj, 291, 747
\bibitem[Wheelock et al.(1994)]{wheelock94} Wheelock, S. L., et al. 1994,
{\it IRAS} Sky Atlas Explanatory Supplement (Publ. 94-11, Pasadena: JPL)
\bibitem[Whitney et al.(2003)]{whitney3}
 Whitney, B. A., Wood, K., Bjorkman, J. E., \& Cohen, M., 2003, \apj, 598, 1079
\bibitem[Wright et al.(2003)]{Tycho2}
 Wright, C. O., Egan, M. P., Kraemer, K. E., \& Price, S. D., 2003, \aj, 125, 359
%\textcolor[named]{Red}{
%\bibitem[Voges et al. 2000]{ROSAT}
%Voges W., Aschenbach B., Boller Th., Brauninger H., Briel U., Burkert W.,
%Dennerl K., Englhauser J., Gruber R., Haberl F., Hartner G., Hasinger G.,
%Pfeffermann E., Pietsch W., Predehl P., Schmitt J., Trumper J., Zimmermann U.
%2000, {\it IAU Circ.}, 7432, 1,
%{\it ROSAT All-Sky Survey Faint source Catalogue (RASS-FSC)}
%}

\end{thebibliography}
\end{document}